\documentclass[epj]{svjour}

\usepackage{graphics}
\usepackage{graphicx}
\usepackage{amsmath}
\usepackage{xcolor}
\usepackage[labelfont=footnotesize, textfont=footnotesize]{caption}
\newcommand{\pol}[1]{\mathaccent"017E{#1}}
\begin{document}
\title{A comprehensive analysis of differential cross sections and analyzing powers in the proton-deuteron break-up channel at 135 MeV\thanks{Supplementary material in the form of a pdf file available from the journal web page at http://*****} }

\titlerunning{A comprehensive analysis of differential cross sections and analyzing powers}

\author{H.~Tavakoli-Zaniani\inst{1,2}\thanks{h.tavakoli.zaniani@rug.nl} \and 
M.~Eslami-Kalantari\inst{2}\thanks{meslami@yazd.ac.ir}\and 
H. R. Amir-Ahmadi\inst{1}\and
M.~T.~Bayat\inst{1}\and  
A.~Deltuva\inst{3}\and
J.~Golak\inst{4}\and 
N.~Kalantar-Nayestanaki\inst{1}\and
 St.~Kistryn\inst{5}\and 
 A.~Kozela\inst{6}\and 
 H. Mardanpour\inst{1}\and
 J.~G.~Messchendorp\inst{1}\thanks{j.g.messchendorp@rug.nl}\and 
 M.~Mohammadi-Dadkan\inst{1,7}\and
 A. Ramazani-Moghaddam-Arani\inst{8}\and 
 R.~Ramazani-Sharifabadi\inst{1,9}\and 
 R.~Skibi{\'{n}}ski\inst{4}\and 
 E.~Stephan\inst{10}\and
 H.~Wita{\l}a\inst{4}
}

%
%
\institute{KVI-CART, University of Groningen, Groningen, the Netherlands \and
 Department of Physics, School of Science, Yazd University, Yazd, Iran \and 
 Institute of Theoretical Physics and Astronomy, Vilnius University, Lithuania \and 
 M.Smoluchowski Institute of Physics, Jagiellonian University, Krak$\acute{o}$w, Poland \and
 Institute of Physics, Jagiellonian University, Krak$\acute{o}$w, Poland \and 
 Institute of Nuclear Physics, PAS, Krak$\acute{o}$w, Poland \and 
 Department of Physics, University of Sistan and Baluchestan, Zahedan, Iran \and 
 Department of Physics, Faculty of Science, University of Kashan, Kashan, Iran \and
 Department of Physics, University of Tehran, Tehran, Iran \and 
 Institute of Physics, University of Silesia, Chorz$\acute{o}$w, Poland}

%
\date{Received: date / Revised version: date}
%
\abstract{
A selection of measured cross sections and vector analyzing powers, $A_{x}$ and $A_{y}$, are presented for the $\pol{p}{d}$ break-up reaction. The data are taken with a polarized proton beam energy of 135~MeV using the Big Instrument for Nuclear-polarization Analysis (BINA) at KVI, the Netherlands. With this setup, $A_{x}$ is extracted for the first time for a large range of energies as well as polar and azimuthal angles of the two outgoing protons. For most of the configurations, the results at small and large relative azimuthal angles differ in behavior when comparing experimental data with the theoretical calculations. We also performed a more global comparison of our data with theoretical calculations using a chi-square ($\chi^{2}$) analysis. The cross-section results show huge values of $\chi^{2}$/d.o.f.. The absolute values of $\chi^{2}$/d.o.f. for the components of vector analyzing powers, $A_{x}$ and $A_{y}$, are smaller than the ones for the cross section, partly due to larger uncertainties for these observables. However, also for these observables no satisfactory agreement is found for all angular combinations. This implies that the present models of a three-nucleon force are not able to provide a satisfactory description of experimental data. 
} 
\maketitle

\section{Introduction}
\label{intro}
\begin{figure*}
\centering
\includegraphics[scale=0.6]{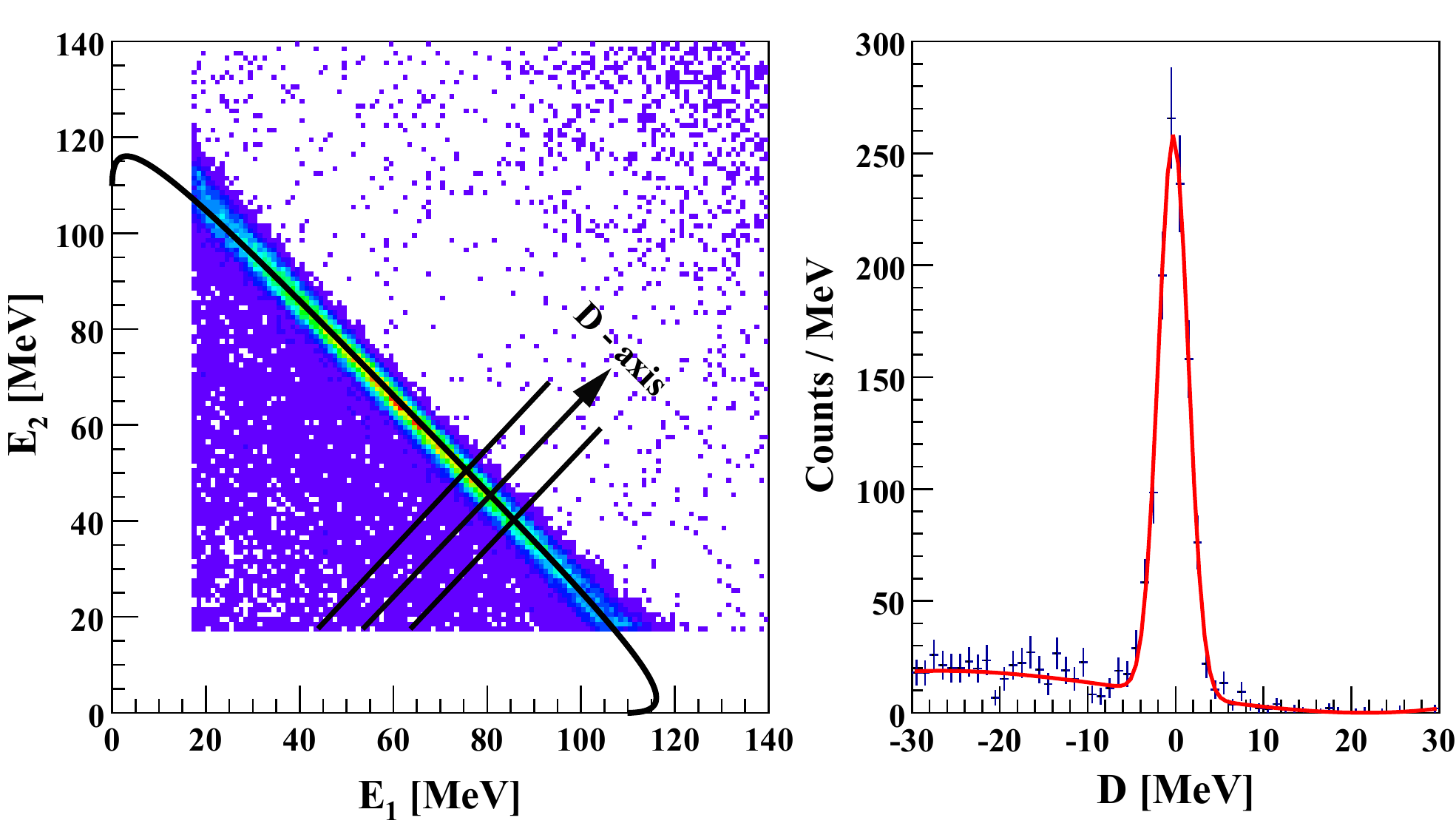} 
\caption{The left panel shows $E_{2}$ versus $E_{1}$, energies of the two outgoing protons at ($\theta_{1}$, $\theta_{2}$, $\phi_{12})=(24^{\circ}\pm2^{\circ}$, $24^{\circ}\pm2^{\circ}$, $180^{\circ}\pm5^{\circ}$). The solid line shows the kinematical $S$-curve calculated for the central values of the experimental angular ranges. The right panel is the projection of events along the D-axis for one slice shown in the left panel.}
\label{fig:1}
\end{figure*}

Although the nucleon-nucleon (2N) interaction has been studied extensively in the past using proton-proton and proton-neutron scattering data, the role of higher-order forces, such as the three-nucleon force (3NF) remains mysterious. The need for an additional three-nucleon potential became evident when comparing three-body scattering observables~\cite{kuro2002,k2002} and binding energies of light nuclei with state-of-the-art calculations~\cite{pieper01}. The two nucleon force models such as CD Bonn, Argonn V18, Reid93, Nijmegen I and Nij\-me\-gen II~\cite{Machleidt2001,Wiringa1995,Stoks1993} are able to describe the two nucleon systems very well below the pion-production threshold. The next step would be to significantly extend the world database in the three-nucleon scattering system as a benchmark to eventually have a better understanding of the structure of the three-nucleon interaction. For almost all observables in nucleon-deuteron elastic scattering, the calculations which only include two nucleon forces (2NFs) fail to a large extent to describe the data, in particular at energies above 60 MeV and at large center-of-mass scattering angle. In addition to the elastic channel, the deuteron break-up reaction offers rich spectrum of kinematical configurations and as such provides a good testing ground for understanding the structure of the nuclear force~\cite{Nasser2012}. Several theoretical approaches have been developed, such as a dynamic $\Delta$-isobar~\cite{deltuva03II} and the Tucson-Melbourne~\cite{witala} 3NFs and these have been embedded within rigorous calculations using the Faddeev-type equations by, for example, Bochum-Krak$\acute{\rm{o}}$w~\cite{witala,witala1988,witala98,witala01} and Hannover-Lisbon~\cite{deltuva03II,deltuva03I,Deltuva2015,Deltuva2008} groups. Besides these phenomenological approaches, also two- and three-nucleon for\-ces 3NF have been constructed from chiral perturbation theory (ChPT). The leading 3NF in ChPT shows significant contributions to the nuclear force~\cite{Machleidt2017}, but the most advanced nowadays complete and consist chiral calculations at the third order of chiral expansion~\cite{Epelbaum2020} deliver the 3NF data description of the similar quality to the one from semi-phenomenological models. 

The proton-deuteron break-up is a suitable reaction to study three nucleon systems one can measure various observables in a large part of the available phase space of this reaction. In this paper, the cross sections and vector analyzing powers, $A_{x}$ and $A_{y}$, for $d(\pol{p}, pp)n$ reaction at 135~MeV are extracted from configurations where the two final-state protons scatter at small polar angles between $14^{\circ}$-$30^{\circ}$ . The data taken at other scattering angles have been reported in Ref.~\cite{bayat2020,bayat2020fb,bayat2019Eu}.

\section{Experimental setup}
\label{Exp}
The $\pol{p}{d}$ break-up reaction was studied using a polarized proton beam of 135 MeV impinging on a liquid deuterium target which was located at the center of BINA (Big Instrument for Nuclear-polarization Analysis).
The po\-la\-rized beam is provided with POLIS (POLarized Ion Sou\-rce)~\cite{FRI95}. The beams of (polarized) protons and deuterons are accelerated by AGOR (Accelerateur Groningen ORsay)~\cite{AGOR} at KVI, the Netherlands. The proton-deuteron break-up reaction was studied with BINA. The BINA detector is particularly suited to study the elastic and break-up reactions at intermediate energies. BINA is composed of two major parts, the forward-wall and the backward-ball. 
The forward-wall measures the energy and scattering angles of final-state particles in the range $10^{\circ}$-$37^{\circ}$. The forward-wall is composed of three main parts, Energy scintillators ($E$-scintillators), $\Delta E$-scintillators, and a Multi-Wire Proportional Chamber (MWPC).
The backward-ball is made of 149 small cut pyramid-shaped scintillator detectors by a ball-shaped detector which covers the rest of the polar angles up to $165^{\circ}$. 
Therefore, the BINA detector covers almost the complete phase space of the break-up and elastic reactions. For a more detailed description of the detector, we refer to~\cite{Hajar2020,hajarthesis}. In this work, we present the results of break-up configurations in which the final-state protons are registered in coincidence by the forward-wall.

\section{Data analysis}
\label{Analysis}
The data analysis of $\pol{p}{d}$ break-up reaction, taken with a proton-beam energy of 135~MeV, was performed with the goal of measuring the vector analyzing powers, $A_{x}$ and $A_{y}$, and the differential cross sections.

Events of the break-up reaction are identified by reconstructing the scattering angles and energies of the two final-state protons. During data taking, a hardware trigger was used requiring at least two of the ten $E$-scintillators to give a signal above the threshold ($\sim$1~MeV). These events were further processed offline by combining the information of the MWPC with the corresponding $E$-scintillators. In this way, two proton candidate tracks were re\-cons\-truc\-ted for further analysis. The $E$-scintillators were calibrated by matching their raw charge-to-digital converter (QDC) information with the expected energy correlation of break-up events. Details of the analysis can be found in Ref.~\cite{hajarthesis}.

The energy correlation between the two outgoing protons, $E_{2}$ versus $E_{1}$, after the calibration for a particular configuration $(\theta_{1}, \theta_{2}, \phi_{12})=(24^{\circ}\pm2^{\circ}, 24^{\circ}\pm2^{\circ}, 180^{\circ}\pm5^{\circ})$ is shown in the left panel of Fig.~\ref{fig:1}, whereby $\theta_{1}$ and $\theta_{2}$ are the polar angles of two outgoing protons and $\phi_{12}$ is their relative azimuthal opening angle.
 The solid line shows the kinematical $S$-curve calculated for the central values of the angular bins. The kinematic variable $S$ corresponds to the arc-length along the kinematic curve with $S=0$ at the point where $E_{1}$ is at its minimum. To measure the break-up observables, at the first step, we make several slices along the kinematical $S$-curve with a window of $\sim$9.5~MeV. We note that the energy resolution, $\sim$4~MeV, is signficantly smaller than this window size. The projection of the indicated region on the line perpendicular to the $S$-curve (D-axis) is shown in the right panel of Fig.~\ref{fig:1}. 
The peak around zero corresponds to break-up events. Most of the events on the left-hand side of the peak are also due to break-up events. In these cases, the protons have lost energy due to hadronic interactions inside the detector. The amount of accidental background is small as can been seen from the small amount of events on the right-hand side of the peak. We fit this spectrum by using a third-order polynomial, representing the hadronic interactions and the accidental background, and a Gaussian function, representing the signal. The extracted number of signal events was corrected by the data-acquisition dead-time and the down-scaling factor. This number is subsequently used to measure  the cross sections and vector analyzing powers.   
 
 The cross section of the break-up reaction can be obtained by: 

\begin{equation}
 \frac{d^5\sigma}{d\Omega_{1}d\Omega_{2}dS}=\frac{N} 
 {Qt\epsilon\Delta\Omega_{1}\Delta\Omega_{2}\Delta S}, \\
\label{eq:1}
\end{equation}
where $N$ is, the number of break-up events in each slice along the $S$-curve corrected for the down-scaling factor and the dead-time, $Q$ is the total integrated charge, $t$ is the number of the scattering centers, $\epsilon$ is the multiplication of all the efficiencies including the MWPC efficiency, hadronic correction and geometrical efficiency, $\Delta\Omega$s are the solid angles for the two outgoing protons and $\Delta S$ is, the width of the selected window in each slice along $S$-curve~\cite{hajarthesis}. 
We studied various sources that we identified as the main contributors to the systematic uncertainty in the cross section measurements. In the following, we briefly summarize each of them and we give a description on how magnitudes of corresponding errors have been estimated.

The first source we identified as a contributor to the systematic error is related to uncertainties in the determination of the effective target thickness. Taking into account the bulging of the target, we estimated an effective target thickness of $3.85\pm0.20$~mm. The resulting error in this measurement (5\%) is assigned as a systematic error in the cross section measurements. This value has been estimated by earlier cross section studies of the elastic proton-proton scattering process using similar targets by comparing data with precision calculations of this reaction~\cite{Huismanphd}.
 
The second systematic uncertainty that we considered is related to the error in estimating the fraction of events that suffered from a hadronic interaction in the scintillators of BINA. Since in the calculation of the number of break-up events, we only account for those events for which the energy of both protons are well reconstructed, one needs to correct for the hadronic interaction effect. To determine this effect, we used Monte Carlo studies that are based on the interaction models provided by the GEANT-3 simulation package~\cite{1610988}. Typically, we found that about 12\% of all break-up events suffered from hadronic interactions. The uncertainty of this value (6\%) is assigned as a source of systematic uncertainty. It has been estimated by taking the difference between the number of hadronic background events derived from simulations with the value estimated from a fit of the measured D spectrum (right panel of Fig.~\ref{fig:1})~\cite{ahmad2011,ahmadphd,Mardanpourphd}. 

The third source of systematic uncertainty is associated with the trigger efficiency. This efficiency has been studied using Monte Carlo simulations based on GEANT-3. It was found that for break-up events whereby $\phi_{12}$ is larger than 20$^\circ$, the trigger efficiency is about 98\% and that it drops to 88\% for selected events associated with $\phi_{12}$=20$^\circ$. To be conservative, we assigned a systematic error due to the trigger efficiency by taking the observed inefficiencies using the Monte Carlo results, therefore 2\% for $\phi_{12}>20^\circ$ and 12\% for $\phi_{12}=20^\circ$~\cite{Eslamiphd}. 

The fourth source of systematic error is due to uncertainties in the efficiency determination of the MWPC. Proton tracks from the elastic proton-deuteron scattering process were identified using the information of the $E$ and $\Delta$$E$ detectors. The $E-\Delta$$E$ hodoscope provides a grid that is used to map onto the MWPC. This allows us to measure the MWPC efficiency for protons at various locations corresponding to every $E-\Delta$$E$ hodoscopy. Typically, we found an efficiency of about (92$\pm$1)\% for each proton, whereby the error corresponds to statistical fluctuations of the unbiased data sample that is used in this study. We associated a systematic error due to uncertainties of the MWPC efficiency for the cross section measurements by summing up the efficiency errors of the two final-state protons, {\it i.e.} 2\%~\cite{Eslamiphd}.

The total systematic uncertainties for the cross sections at small relative azimuthal angles ($\leq20^{\circ}$) are about 14\% and for the larger relative azimuthal angles ($> 20^{\circ}$) is about 9\%. For this, we added up, quadratically, the systematic errors of the various sources assuming them to be independent.

To measure the vector analyzing powers, the number of break-up events were normalized to the collected beam charge for the two polarization states (up and down). The relation between the normalized number of events with the polarized beam, $N_{\xi,\phi_{12}}^{s}(\phi)$, and unpolarized beam, $N_{\xi,\phi_{12}}^{0}$, is given by~\cite{Ohlsen1981}:

\begin{eqnarray}
N_{\xi ,\phi_{12}}^{s}(\phi)=&&N_{\xi ,\phi_{12}}^{0}(1+p_{z}^{s}A_{y}(\xi ,\phi_{12})\cos\phi \nonumber\\
&&-p_{z}^{s}A_{x}(\xi ,\phi_{12})\sin\phi),
\label{eq:2}
\end{eqnarray}
where $s$ indicates the spin of the beam and $\xi$ defines a given kinematical point ($\theta_{1}$, $\theta_{2}$, $S$). The component of the vector polarization of the beam is given by $p_z$ and the vector analyzing powers are indicated by $A_{x}$ and $A_{y}$. Here, $\phi$ is the angle between
quantization axis for the polarization and the normal to the scattering plane of the first nucleon in the laboratory frame of reference, with $\phi_{1}=0$. 
Since the statistics obtained with an unpolarized beam was limited, we extracted the spin observables by solely using $N_{\xi, \phi_{12}}^{{\uparrow}}(\phi)$ and $N_{\xi, \phi_{12}}^{{\downarrow}}(\phi)$, corresponding to the normalized number of events for the spin-up and spin-down polarized beams, respectively. The analyzing powers $A_{x}$ and $A_{y}$ are extracted using the following relation:

\begin{equation}
\begin{split}
 f_{\xi ,\phi_{12}}(\phi) & = \frac{N_{\xi ,\phi_{12}}^{{\uparrow}}(\phi)-N_{\xi ,\phi_{12}}^{{\downarrow}}(\phi)}{N_{\xi ,\phi_{12}}^{{\uparrow}}(\phi)p_{z}^{\downarrow}-N_{\xi ,\phi_{12}}^{{\downarrow}}(\phi)p_{z}^{\uparrow}} \\ 
 & =A_{y}(\xi ,\phi_{12})\cos\phi-A_{x}(\xi ,\phi_{12})\sin\phi,
\end{split}
\label{eq:3}
\end{equation}
where $p_{z}^{\uparrow}$ and $p_{z}^{\downarrow}$ are the values of up ($0.57\pm0.03$) and down ($-0.70\pm0.04$) beam polarizations. The polarization of the proton beam is defined as:

\begin{equation}
p_{z} = \frac{N^{+}-N^{-}}{N^{+}+N^{-}},
\end{equation}
where $N^{+,-}$ are the number of particles with a particular spin (up or down). The beam polarization has been determined using the in-Beam Polarimeter (IBP)~\cite{Bieber01} that was installed at the high-energy beam at KVI. The IBP measured regularly the beam polarization by recording the azimuthal asymmetries of the H(\vec p, pp) reaction. The vector analyzing power of the proton-proton scattering process was used as input to the polarization measurements and its uncertainty is the main source of error.
Parity conservation imposes the following restrictions on the components of the vector analyzing powers~\cite{Ohlsen1981}:

\begin{equation}
 \begin{aligned}
 A_{x}(\xi ,-\phi_{12})= -A_{x}(\xi ,\phi_{12}); \\
 A_{y}(\xi ,-\phi_{12})= A_{y}(\xi ,\phi_{12}), 
\label{eq:4}
 \end{aligned}
\end{equation}
where for $\phi_{12}=180^{\circ}$, we expect $A_{x}=0$. By taking the sum and difference of $f_{\xi, \phi_{12}}(\phi)$ and $f_{\xi,-\phi_{12}}(\phi)$ in combination with the results of Eq.~\ref{eq:4}, the following combination of asymmetries for mirror configurations ($\xi$,$\phi_{12}$) and ($\xi$,$-\phi_{12}$) can be obtained~\cite{Hajar2020,Stephan2010,Stephan2013}:

\begin{equation}
 \begin{aligned}
 g_{\xi ,\phi_{12}}(\phi) &= \frac{f_{\xi ,\phi_{12}}(\phi)+f_{\xi ,-\phi_{12}}(\phi)}{2},  \\
              &=A_{y}(\xi ,\phi_{12})\cos\phi;\\
 h_{\xi ,\phi_{12}}(\phi) &= \frac{f_{\xi ,\phi_{12}}(\phi)-f_{\xi ,-\phi_{12}}(\phi)}{2},  \\
              &=-A_{x}(\xi ,\phi_{12})\sin\phi.\\ 
\label{eq:5}
 \end{aligned}
\end{equation}
The components of vector analyzing-power values, $A_{x}$ and $A_{y}$, are obtained from the fits of Eq.~\ref{eq:5} for various kinematical configurations. 

\begin{figure}
\includegraphics[scale=0.4]{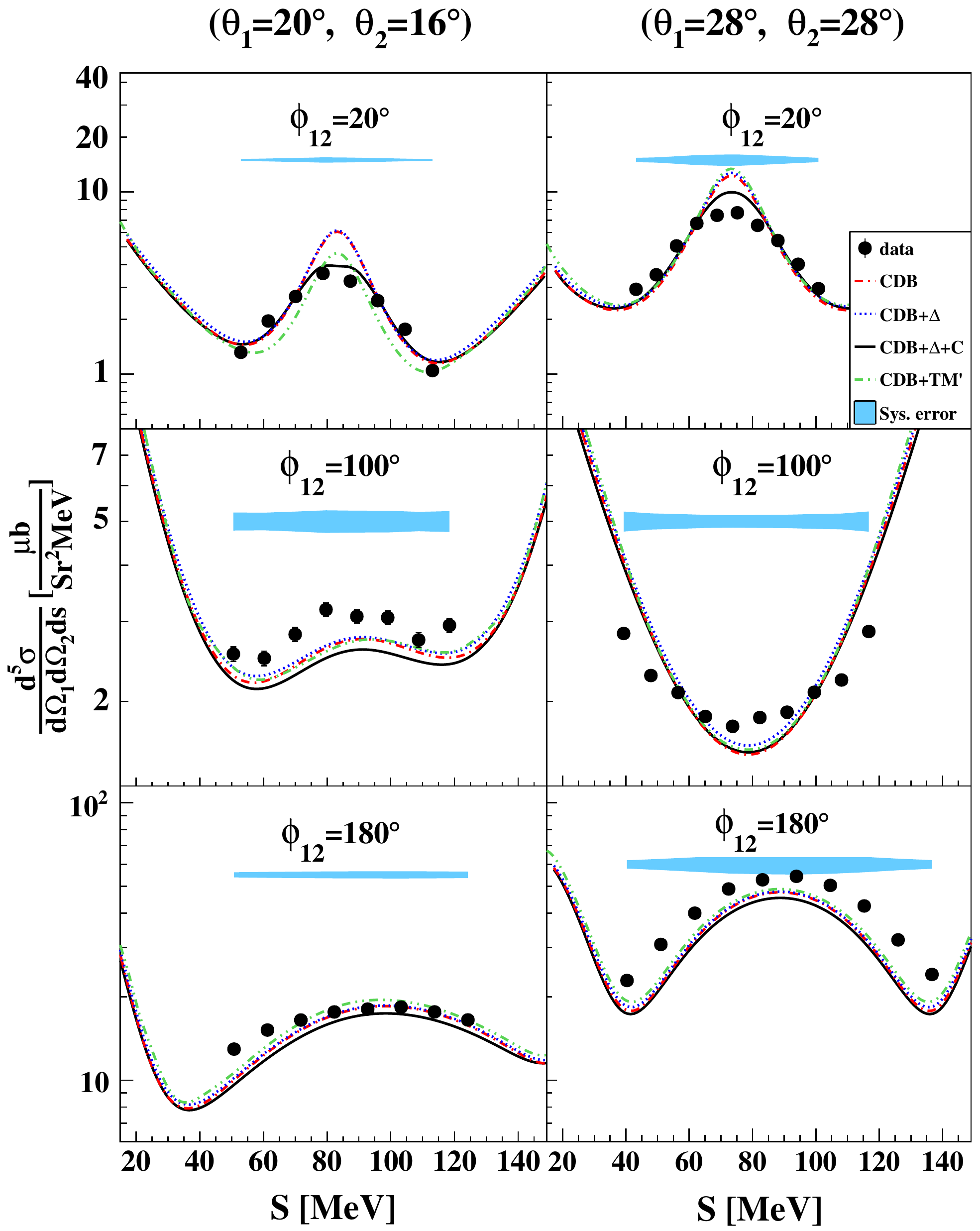}
\caption{\label{fig:2}Cross sections at $(28^{\circ}$,$28^{\circ})$ (left) and $(28^{\circ}$,$24^{\circ})$ (right) as a function of $S$ at small, intermediate and large relative azimuthal angles for data taken with a proton beam of 135~MeV. Error bars show the statistical uncertainties for the data points. The red (dotted-dashed), blue (dotted), black (solid) and green (double dotted-dashed) lines show predictions of Faddeev calculations using CD-Bonn, CDB$+\Delta$, and CDB$+\Delta+$Coulomb and CDB+TM$'$ calculations~\cite{witala,witala1988,witala98,witala01,Deltuva2008,Deltuva2005,Deltuva2013,Witala2009}, respectively. The cyan bands depict the systematic uncertainties (2$\sigma$).}
\end{figure}

\begin{figure}
\includegraphics[scale=0.4]{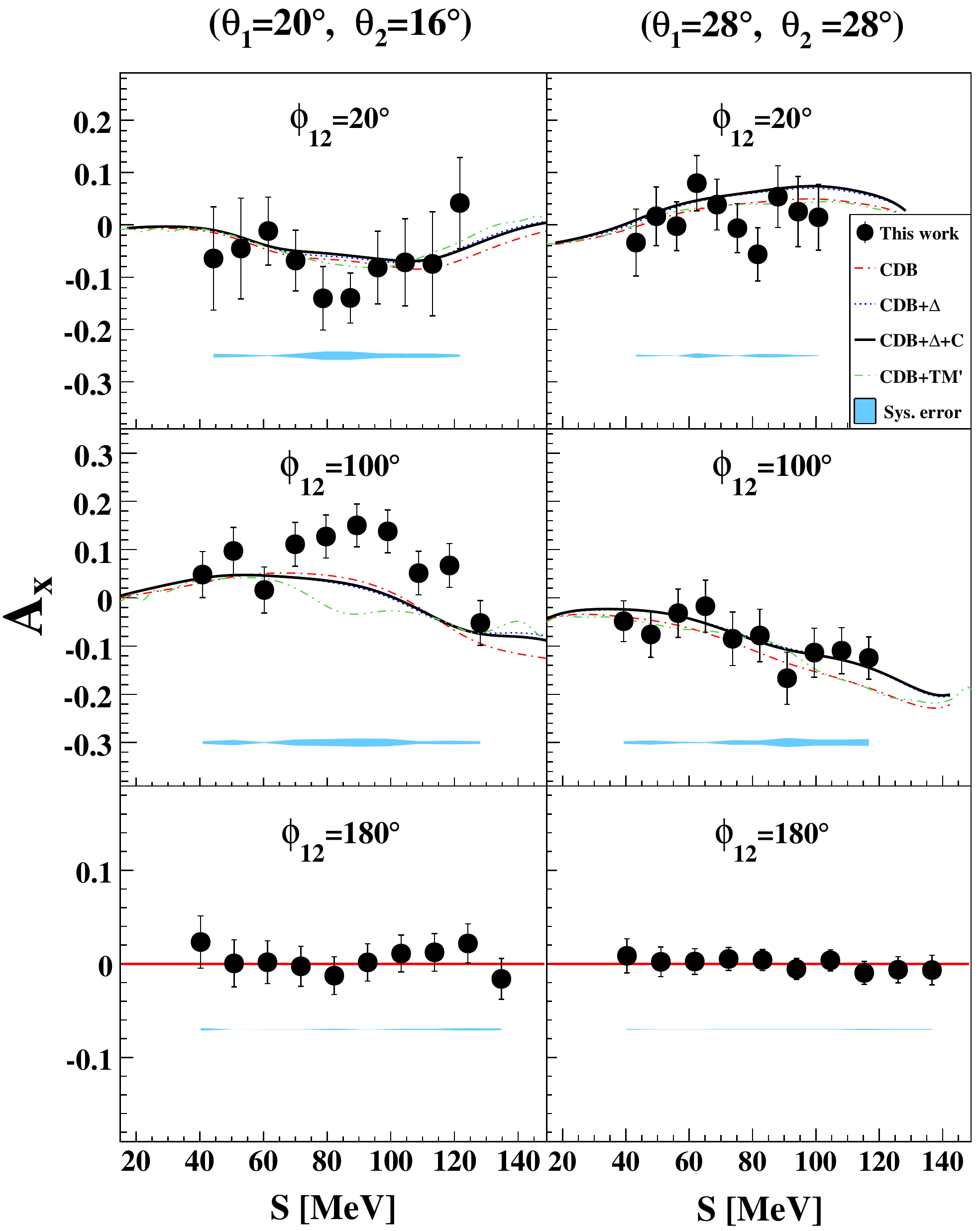}
\caption{\label{fig:3}Same as Fig.~\ref{fig:2} except for $A_{x}$. The red lines in the bottom panels correspond to a zero line.}
\end{figure} 

\begin{figure}
\includegraphics[scale=0.4]{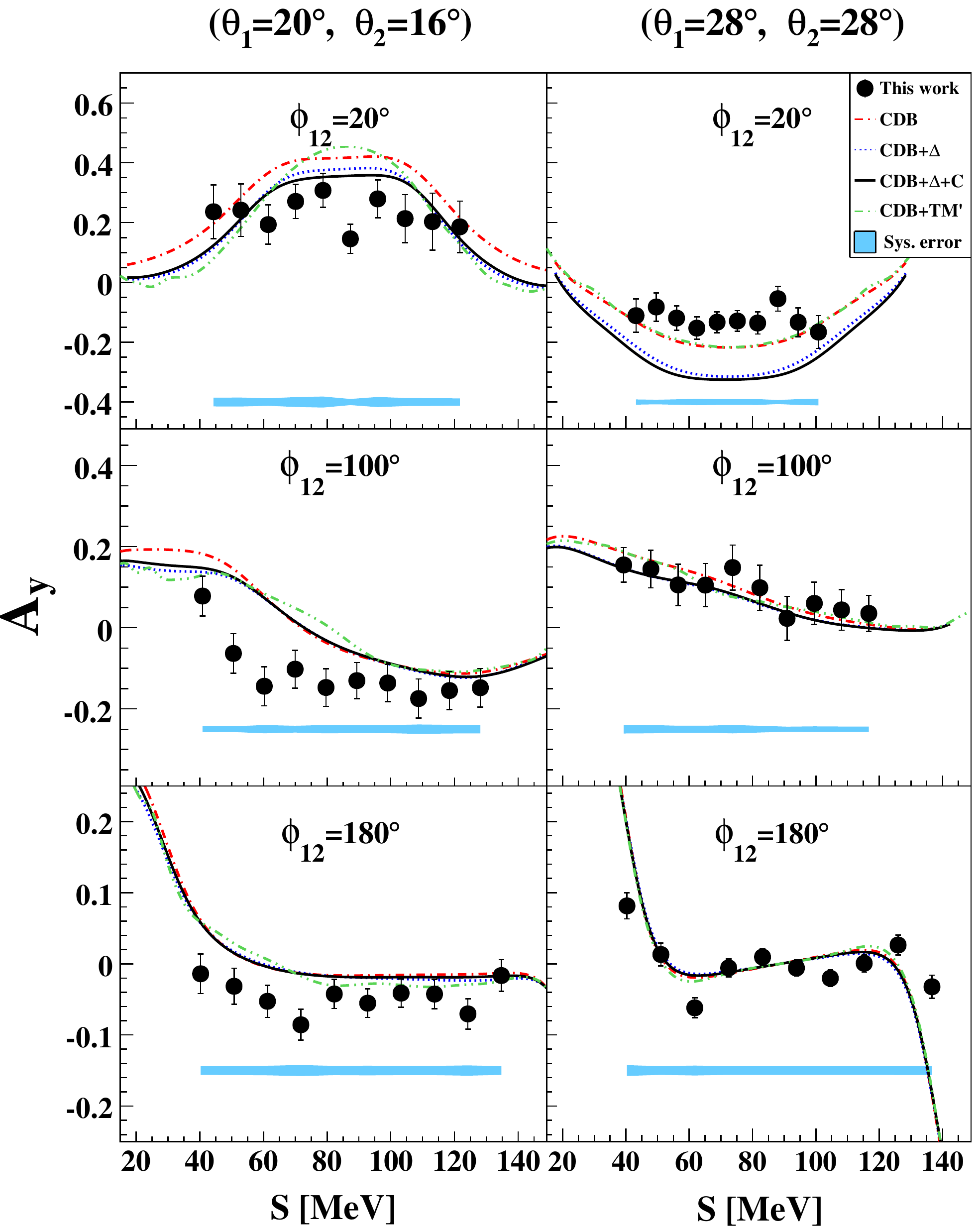}
\caption{\label{fig:4}Same as Fig.~\ref{fig:2} except for $A_{y}$.}
\end{figure}

The error of the beam polarization is about 6\%. For instance, the beam polarization for the down-mode
has been measured at a value of $\sim0.70\pm0.04$, which gives rise to 6\% systematic uncertainty in the beam polarization. We estimated the impact of the polarization uncertainty on the analyzing powers by recalculating both analyzing powers with an input polarization that differs by $+$6\% ($-$6\%) for the spin-up (down) mode. The difference with the results using the nominal values of the beam polarizations is used as an estimate of the corresponding systematic error. We also considered a systematic error due to asymmetries that are induced by rate- or polarization-dependent differences in detection efficiencies that do not cancel in Eq.~\ref{eq:3}. This uncertainty has been estimated by exploiting data at particular kinematical configurations for which the vector analyzing powers are known or constrained. For $A_y$, we have analyzed various symmetric configurations for which both protons scatter to the same polar angle with a relative azimuthal angle of 180$^{\circ}$. By taking the average vector analyzing power for the covered $S$-range, one expects a value of zero. We have performed a fit with a free offset value to the data and we used the corresponding offset as a measure of the systematic uncertainty for $A_y$. To estimate the systematic error for $A_x$, we analyzed the data for a relative azimuthal angle of 180$^{\circ}$ for which $A_x$ should be zero, and extracted the corresponding value for $A_x$ as a function of $S$. Subsequently, these results are fitted with a zeroth-order polynomial and its value is used as an estimate for the corresponding systematic error. This error and the error in the polarization are added in quadrature assuming them to be independent, to form the total systematic uncertainty.

\begin{figure*}
\centering
\includegraphics[scale=0.6]{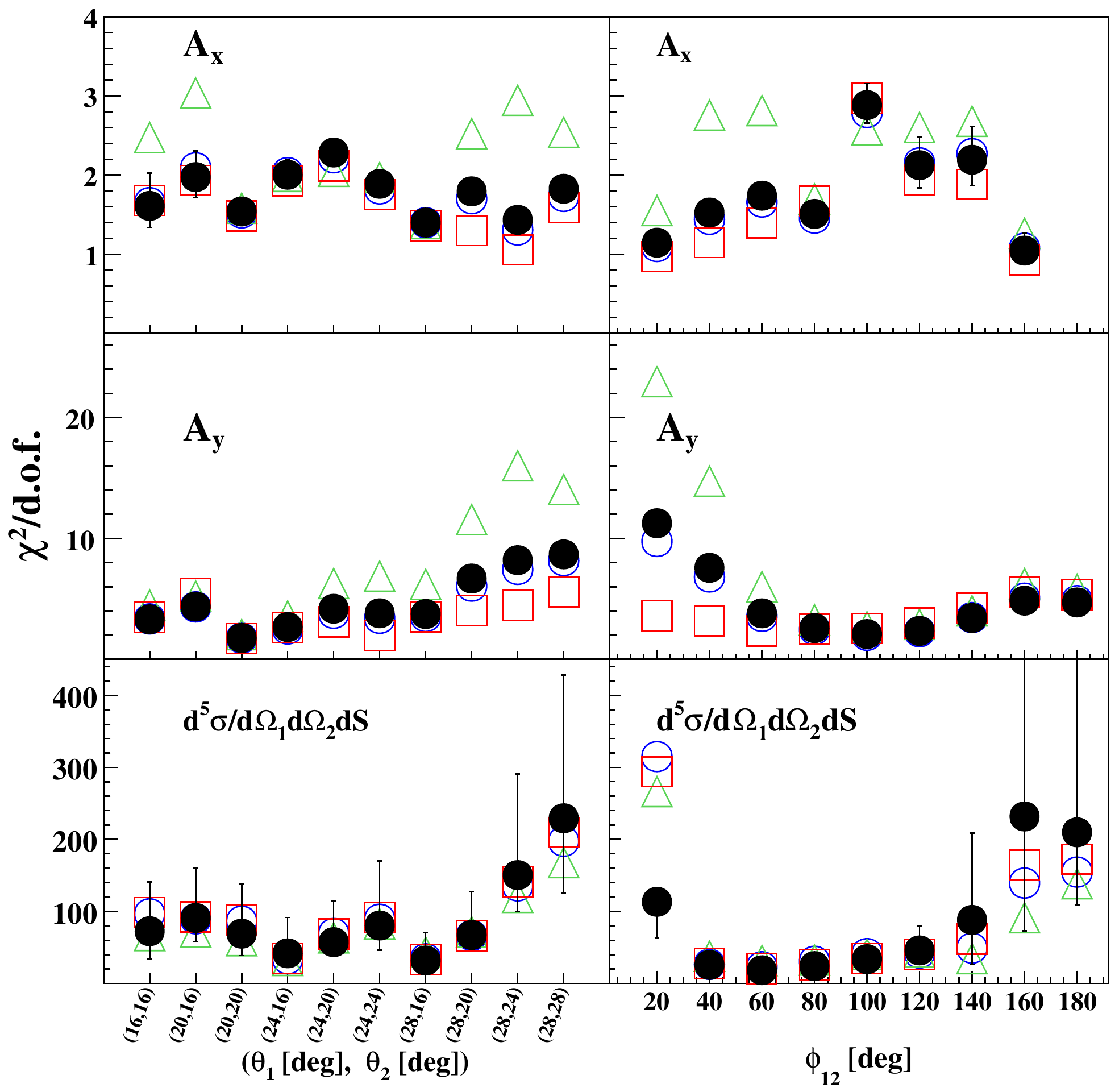}
\caption{\label{fig:5} The results of $\chi^{2}/$d.o.f. versus ($\theta_{1}$ and $\theta_{2}$) (left panels) and versus  $\phi_{12}$ (right panels) for different observables ($A_{x}$, $A_{y}$ and $d^{5}\sigma/d\Omega_{1}d\Omega_{2}dS$) for data taken with a proton-beam energy of 135~MeV. The symbols show the theoretical calculations such as CDB (squares), CDB$+\Delta$ (open circles) and CDB$+\Delta+$Coulomb (solid circles) and CDB+TM$'$ (triangles). The error bars reflect the systematic uncertainty. For details, see text.}
\end{figure*}

\begin{figure*}
\centering
\includegraphics[scale=0.6]{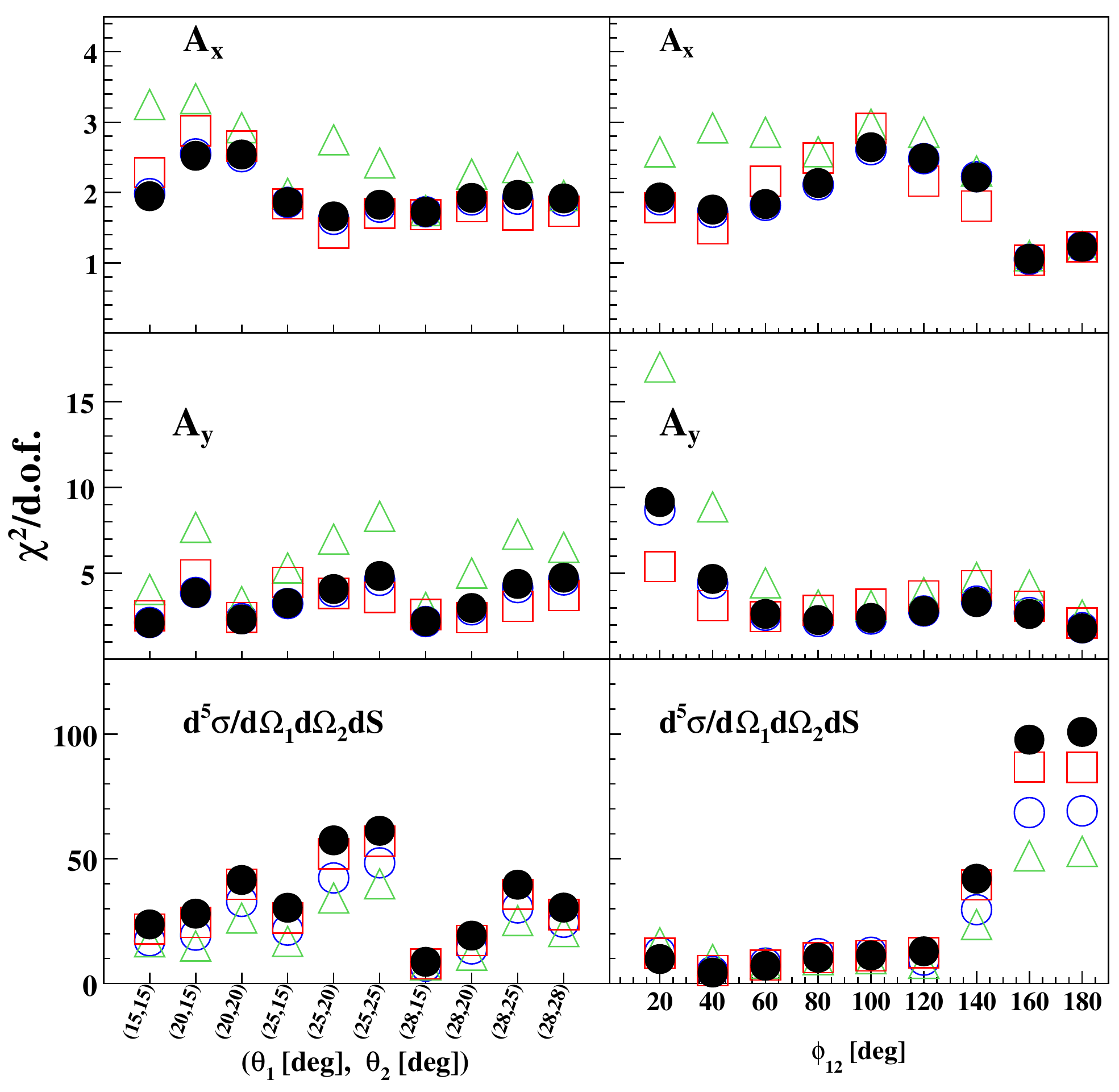}
\caption{\label{fig:6}Same as Fig.~\ref{fig:5} except for 190~MeV.}
\end{figure*}

Figures~\ref{fig:2}--\ref{fig:4} show the cross-sections and vector analyzing powers as a function of $S$ for symmetric and asymmetric configurations at small, intermediate and large relative azimuthal angles. The results of our analysis are indicated as black dots. The error bars indicate statistical uncertainties which are in some cases smaller than the symbol sizes. The cyan bands depict the systematical uncertainty whereby the width corresponds to 2$\sigma$. The various lines present the results of Faddeev calculations using 2NF and 2N+3NF models. The results show a different behavior between the data and theoretical calculations at small and large relative azimuthal angles. 

The results of the cross sections and vector analyzing powers as a function of $S$ for about hundred configurations (with $14^{\circ}<\theta_1<30^{\circ}$, $14^{\circ}<\theta_2<30^{\circ}$ and $0^{\circ}<\phi_{12}<180^{\circ}$) for incident proton energy of 135 MeV were extracted. A small subset of vector analyzing power data for selected symmetric configurations was presented in Ref.~\cite{Hajar2020}.
An extensive overview of all the results can be found in the supplementary material associated with this paper~\cite{Ref}. In general, we observe similar patterns for other scattering angles comparable to the ones shown in Figs.~\ref{fig:2}--\ref{fig:4} with respect to small and large relative azimuthal angles. Similar behaviours were also observed in the measurements of the same observables at 190 MeV~\cite{maisam2020}.

To have a more efficient study and to compare globally the theoretical predictions with the complete data set, a global analysis is performed with averages of observables~\cite{hajarthesis,bayat2020}. In this paper, we compare our data with the theoretical predictions using a chi-square analysis for several ($\theta_{1}$, $\theta_{2}$) and as a function of $\phi_{12}$ as another approach to perform a global analysis. The quantity $\chi^2$ per degree of freedom is defined by 

\begin{equation}
\chi_{m}^{2}/d.o.f.=\frac{1}{N-1} \sum\limits_{i=1}^{N}\Big\{ \frac{O_{i}-T_{i}^{m}}{\sigma_{i}}\Big\} ^2,
\label{eq:3-5}
\end{equation}
where $N$ is the number of specific configuration in ($S$, $\theta_{1}$, $\theta_{2}$, $\phi_{12}$), $O_{i}$ is one of the observables ($A_{x}$, $A_{y}$, or $d^{5}\sigma/d\Omega_{1}d\Omega_{2}dS$), $\sigma_{i}$ is the measured statistical error of a data point and $T_{i}^{m}$ is the results of the theoretical calculation whereby the sup-index $m$ refers to a specific model, namely, CDB, CDB+TM$'$, CDB$+\Delta$ and CDB$+\Delta+$Cou\-lo\-mb, the sub-index $i$ refers a specific configuration in ($S$, $\theta_{1}$, $\theta_{2}$, $\phi_{12}$) and $N$ is the number of specific configuration in ($S$, $\theta_{1}$, $\theta_{2}$) in the right panels of Figures~\ref{fig:5} and~\ref{fig:6} or the number of specific configuration in ($S$, $\phi_{12}$) in the left panels of Figures~\ref{fig:5} and~\ref{fig:6}. 
Figures~\ref{fig:5} and~\ref{fig:6} show the results of $\chi^{2}$/d.o.f. for sum over ($S$, $\phi_{12}$) for specific ($\theta_{1}$, $\theta_{2}$) versus the angular combination ($\theta_{1}$ and $\theta_{2}$) (left panels) and for sum over ($S$, $\theta_{1}$, $\theta_{2}$) for specific $\phi_{12}$ versus $\phi_{12}$ (right panels) for different observables ($A_{x}$, $A_{y}$ and $d^{5}\sigma/d\Omega_{1}d\Omega_{2}dS$) for data taken with a proton-beam energy of 135~MeV and 190~MeV~\cite{Maisam}, respectively. The asymmetric error bars reflect the systematic uncertainty of the data with respect to one of the theoretical calculations, namely CDB$+\Delta+$Coulomb. These errors were obtained by adding and subtracting the estimated total systematic error to and from the data resulting in two alternative chi-square values corresponding to the edges of the error bars. 

\section{Discussion}
\label{Dis}
As observed in Fig.~\ref{fig:2}, at small azimuthal opening angles the results are closer to the predictions of the theoretical approach that deploy CDB$+\Delta+$Coulomb potential. This demonstrates that the Coulomb effect is sizeable for this observable at these configurations. Note that in this case the relative energy between the two protons is small. For large relative azimuthal angles, the model based on the CDB+TM$'$ potential appears to be the closest to the experimental data, albeit the differences between the various models are in general small. For intermediate values of $\phi_{12}$, the predicted shape of the cross sections differ significantly with the data. In Figs.~\ref{fig:3} and~\ref{fig:4}, the measurements of $A_{x}$ for relative azimuthal angles of $180^{\circ}$ is found to be consistent with zero as expected from Eq.~\ref{eq:4}. This demonstrates that our procedure to extract the analyzing powers does not suffer from experimental asymmetries. This is also confirmed by our estimate of the systematic uncertainty, which is found to be small. Our polarization observables are reasonably well described by the calculations for kinematical configurations at which the three-nucleon force effect is predicted to be small. However, striking discrepancies are observed at specific configurations, in particular in cases where the relative azimuthal angle between the two outgoing protons becomes small. In this range, the measured values of $A_{y}$ is close to the results of the 2NF calculation. Although, the disagreement is still significant, the effects of the Coulomb force are very small. The addition of the TM$'$3NF makes the agreement even worse. Therefore, the origin of this discrepancy must lie in the treatment of 3NFs. The same behavior was observed for the data taken at a beam energy of 190 MeV~\cite{Maisam}. Possibly, the modeling of short-range 3NF must be significantly improved as $\it{e.g.}$ chiral perturbation theory, for which the data presented in this paper and those in Ref.~\cite{Maisam} can be used as a benchmark.

The results of the analyzing powers for different combinations of ($\theta_{1}=\theta_{2}$, $\phi_{12}$) for small $\phi_{12}$ which corresponds to $d(\pol{p},{\rm{^{2}He}})n$ for 135~MeV proton beam energy were also compared to the results using a proton beam with an energy of 190~MeV~\cite{Mardanpour2010} to study the spin-isospin sensitivity of the 3NF models~\cite{Hajar2020,Hajar2020fb,Hajar2019Eu}. 

By inspecting Fig.~\ref{fig:5}, we note that the absolute values of the $\chi^{2}/$d.o.f. for the analyzing powers $A_{x}$ and $A_{y}$ appear to behave better than the ones for the cross section. However, also for analyzing powers, there are clear trends to be observed in which all the model predictions deviate, beyond statistical and systematic uncertainties, from the data.
For $A_{y}$, the trend observed in the plots as a function of polar angle combination looks similar to what is observed for the cross section. The models show a larger discrepancy towards larger angles. The trends as a function of $\phi_{12}$ are vastly different compared to the ones observed in the cross section. Although $A_{y}$ features a worse agreement towards small $\phi_{12}$ (and partly large $\phi_{12}$), the observable $A_{x}$ is well predicted at small $\phi_{12}$ except for CDB+TM$'$. By comparing all the model predictions, the calculation based on the CDB$+\Delta+$Coulomb model is the most compatible with the data.

By comparing the results between the two energies, see Figs.~\ref{fig:5} and \ref{fig:6}, in general, similar trends as a function of $\phi_{12}$ are observed for the cross section and $A_{y}$. For the cross section data taken at 190 MeV, the calculation that is based on CDB$+\Delta+$Coulomb potentials shows the worst agreement, in particular for large values of $\phi_{12}$. The sensitivity to 3NF effects appears to be larger at the higher energy. For both energies, it is clear that the inclusion of the TM$'$ 3NF is by far not sufficient to remedy the observed discrepancies. We also note that the Coulomb effect is very small for both spin observables. By globally reviewing the chi-square data, we note that CDB+TM$'$ gives the worst description of the data for analyzing powers.

\section{Summary and conclusions}
\label{summary}
Finding a suitable theory of nuclear forces is one of the main challenges in nuclear physics. 
To study the three nucleon systems, the reaction $d(\pol{p}, pp)n$ was studied at KVI using a polarized proton beam.
In this paper, the results of the vector analyzing powers, $A_{x}$ and $A_{y}$, and the cross section for data taken with a proton-beam energy of 135 MeV are presented. Moreover, we performed a global review of a rich set of cross section and vector analyzing-power data taken with proton-beam energies of 135 MeV and 190~MeV. The results were compared with theoretical Faddeev calculations using 2N and 2N+3NF models such as CD-Bonn, CDB$+\Delta$, CDB$+\Delta+$Coulomb and CDB$+$TM$'$~\cite{witala,witala1988,witala98,witala01,Deltuva2008,Deltuva2005,Deltuva2013,Witala2009} for the kinematics in which both protons scatter to polar angles  smaller than $30^{\circ}$ and with a relative azimuthal opening angle varying between $20^{\circ}$ and $180^{\circ}$. The results of cross sections and analyzing powers, $A_{x}$ ans $A_{y}$, as a function of $S$ for different configurations ($\theta_{1}$, $\theta_{2}$, $\phi_{12}$) are shown in Figs.~\ref{fig:2}--\ref{fig:4}. At small azimuthal opening angles, the calculation, which is based on the extended CDB+$\Delta$ and with Coulomb corrections, CDB+$\Delta+$Coulomb, shows a smaller discrepancy with the data than the other calculations. The results show that there is a general disagreement between the data and the calculations including a 3NF. In particular, predictions for the vector analyzing powers show a systematic deficiency at small relative azimuthal angles, which corresponds to small relative energies. In this range, the data for $A_{y}$ is closest to the three-body calculation that is based on a 2N potential. The addition of 3NF makes the agreement even worse. 

The results of the global review show very large values of $\chi^{2}$/d.o.f. for the cross sections at specific scattering and relative azimuthal angles. The deviations, independent of the model and beam energy, appear to increase towards large values of $\phi_{12}$. This implies that the present models are not able to provide a reasonable description of the data. The absolute $\chi^2$/d.o.f. for the analyzing powers $A_{x}$ and $A_{y}$ are much closer to unity than the ones observed for the cross sections. However, also for these observables no satisfactory agreement is found for all the angular combinations ($\theta_{1}$, $\theta_{1}$) and $\phi_{12}$.
%
\section*{Acknowledgement}
The authors acknowledge the work by AGOR cyclotron
and ion-source groups at KVI for delivering the highquality
polarized beam. This work was partly supported by the Polish National Science Centre under Grant No. 2012/05/B/ST2/02556 and 2016/22/M/ST2/00173. The numerical calculations of Bochum-Krakow group were partially performed on the supercomputer cluster of the JSC, J\"ulich, Germany.
%
\bibliographystyle{epj}
\bibliography{Mypaper6}

\end{document}